# Metamimetic Games

## Modeling Metadynamics in Social Cognition

David Chavalarias[1],
Center for Research in Applied Epistemology (CREA),
Ecole Polytechnique, Paris, France,
www.polytechnique.edu

***Abstract:***

Imitation is fundamental in the understanding of social system dynamics. But the diversity of imitation rules employed by modelers proves that the modeling of mimetic processes cannot avoid the traditional problem of endogenization of all the choices, including the one of the mimetic rules. Starting from the remark that human reflexive capacities are the ground for a new class of mimetic rules, I propose a formal framework, metamimetic games, that enable to endogenize the distribution of imitation rules while being human specific. The corresponding concepts of equilibrium - counterfactually stable state - and attractor are introduced. Finally, I give an interpretation of social differentiation in terms of cultural co-evolution among a set of possible motivations, which departs from the traditional view of optimization indexed to criteria that exist prior to the activity of agents.

---

[1] david.chavalarias@polytechnique.edu ; home page : www.chavalarias.com

# I. Social system modeling and the grounding problem

**I.1 An interdisciplinary convergence on imitation**

Recent years have seen a burgeoning of interest in the phenomenon of imitation on the part of researchers in many fields. After the seminal work of Tarde (1890) in sociology and Baldwin (1897) in developmental psychology, Girard (1961) takes imitation as the first foundation stone of his theory of the origin of primitive societies, and Donald (1991) considers that the sophistication of human mimetic skill could have been one of the major evolutionary transitions in hominization.

After three of the finest economists of all time, Smith, Keynes and Hayek gave imitation a central role (Dupuy 2004), experimental studies have been conducted revealing the importance of imitation in decision-making processes (Camerer 2003; Hück et al. 1999; Selten & Ostmann 2001). Nowaday, an increasing number of scholars are considering imitation as a key phenomenon in the study of micro-economic behaviors, especially in business and finance where information and communication technologies spawn a proliferation and speed-up of feedback loops (Day & Huang 1990; Frank 2003; Kirman 1991; Orléan 1998).

Today, the study of imitation is a dynamic field of research in developmental psychology (Meltzoff 2002; Gergely et al. 2002; Zelazo & Lourenco 2003) with bridges toward neurology (Arbib 2000; Chaminade et al. 2002) and the theory of mind (Meltzoff & Gopnik 1993; Pacherie 1998).

Finally, in artificial intelligence and robotic, imitation is now seriously considered as a means of constructing socially intelligent artificial agents (Breazeal et al. 2005; Jansen et al. 2003; Zlatev 2000). Evidence that stems from ethology (Tomasello 1999) about the incapacity of animals to imitate in a human way will certainly not weaken this interest for imitation in the study of human behavior.

It stems from this rich literature that imitation plays a central role in human cognitive development and is fully involved, with other interdependent cognitive processes, in everyday life decision-making processes. These conclusions have had interesting echoes in the literature of socio-economic modeling where an increasing number of modellers incorporate mimetic processes into formal approaches to account for the extremely rich structures observed in human societies. The majority of papers related to social modeling concentrates on the particular social dynamics that mimetic processes generate rather than on the problem of interfacing different mechanisms of decision-making process. However, even at this stage, hard modeling issues arise. In particular, the diversity of imitation rules employed by modelers proves that the introduction of mimetic processes into formal models cannot avoid the traditional problem of the endogenization of all choices, including the choice of the imitation rules. In this paper, we explore the consequences of different types of models in artificial settings and propose to take into account some suggestions of sociology and cognitive science to build a new class of models for social modeling. In the following, we will use the term 'individuals' to refer to human beings and the term 'agents' to refer to formal or artificial entities that are representations for human beings in models or simulations.

**I.2 Which rules are the "good" ones?**

In the literature of social systems modeling, the most frequent types of rules are these: (1) *payoffs-biased imitation,* i.e. imitation of the most successful agents in one's neighborhood (Nowak and May 1992) and (2) *conformism* (Axelrod 1997; Galam 1998; Orléan 1985, 1998, Rogers 1988). Here, *conformism* is the propensity of individuals to adopt some behavior when it has already been adopted by some of their neighbors, the propensity being relative to the frequency of that behavior in the neighborhood. To a lesser extent, other imitation processes have been studied, among which we can mention (3): *non-conformist*[i], the propensity of an individual to adopt the behavior of the minority (Arthur 1994; Edmonds 1999) and (4) *prestige* (Henrich & Gil-White 2001). This list of imitation rules is far from exhaustive, and we should note that even for *conformism* or *payoffs-biased imitation*, several technical definitions have been proposed both deterministic and probabilistic (Nowak et al. 1994). Moreover, it is also possible to propose models that include several imitation rules, as some authors already did (Boyd & Richerson 1985; Henrich & Boyd 1998; Janssen & Jager 1999; Kaniovski et al. 2000; Vriend 2002).

This raises an epistemological question for modelers. Which rule(s) for imitation should be considered given the situation under study?
Let's try to be a little more precise. If we represent schematically the models cited above we can remark that all can be described in terms of hierarchies of rules governing a behavior (cf. table 1. for some examples). The proportion of the rules at a given level is determined by the metarules of the level above. In this representation, rules can be interpreted as dynamics principles acting at the population level (like the replicator dynamics) or as a decision-making rule used at the individual level. The emergence of patterns at the collective level is thus understood as a selection of a particular distribution on the set of possible rules and meta-rules. Now the question is: *How are these distributions selected?*

Some scholars have addressed this question from an evolutionary perspective, assuming that the distribution of imitation rules is shaped by natural selection (see for example Henrich & Boyd 1998 - table *1-d*). This approach assumes that there is a unique selection rule, indexed to fitness that drives the entire system from the top (for example, the conformist transmission rule is often viewed as a rule that has been selected during evolution, humans being wired to be conformist to some extent).

But the slow dynamics of genetic processes seems to be incompatible with the fast evolution observed in human socio-economic systems (Feldman & Laland 1996, Frank 2003, Gould 1987, Gintis 2003) that varies on the scale of a lifespan. The challenge is thus to identify an evolutionary process that could lead individuals (or firms in the case of economic systems) to choose between several possible imitation rules while interacting with their environment. Rather than considering agents whose imitation rules exist prior to their socio-economic activities, we have to imagine agents, whose way of being influenced by others is the result of an ongoing interaction with their social environment.

**Table 1: Schematic representation of models**

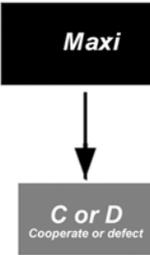

| a.: Nowak and May 1992. | b.: Kaniovski et al. 2000. | c.: Orléan 1998. | d: Henrich and Boyd 1998. | e: Modelers in the social sciences generally represent agents as a hierarchy of rules, where rules at each level evolve under the dynamics defined by their meta-rules. |

Here a strange loop appears: the distribution of imitation rules in a population is the consequence of interactions among agents; the consequences of interactions among agents are defined in terms of their imitation rules. We face a grounding problem concerning the dynamics of imitation rules in the true sense of the term. *Social cognition* can be defined in terms of a collective processing of information that is distributed over all the individuals of a society (Bourgine 2004). These individuals are locally processing information using their rules for decision-making processes. We have thus to study the dynamics of this collective information processing which is a metadynamics relatively to dynamics defined by the agents' rules. Can dynamics and metadynamics coincide? This leads us to reformulate our question in terms close to the notion of *operational closure* (Varela 1983): *Is it possible to endogenize the distribution of decision-making metarules such that this distribution becomes the outcome of the dynamics it defines?*

The point of this article is to demonstrate that there exists at least one class of models that allow us to give a yes answer to this question. We will show on the way that considering the specificity of human imitation, we can formally think of imitation rules as their own metarule.

## I.3 The specificity of human cognition

Another way to address the question of the endogenization of imitation rules will perhaps come from a recent concern in social systems modeling. The complexity of human social systems has no counterpart in other species. For example, considering group coordination, only insect societies, composed of very simple entities, have social structures involving several thousand members. This feature disappears as soon as the repertoire of behavioral possibilities of a species gets wider, and reappears only when it comes to humans (Bourgine 2004, Wilson 1975). But if an anthill is often taken as a metaphor for social systems self-organization, it is clear that the goal for social systems modeling is not to consider humans as cloned insects. What is at stake is rather to find differences between humans and others mammals, which make possible the emergence of highly structured social groups while preserving inter-individual heterogeneity. This has led recently some modelers to propose, as a heuristic for social systems modeling, prioritizing models that could be human specific (Alvard 2003, Bowles & Gintis 2003, Fehr & Fischbacher 2003). In the social sciences, a similar heuristic that particularly focuses on imitation, was formulated a few decades ago by René Girard[ii] (1978):

> *In order to develop a science of man, we must compare human imitation with animal mimetism and separate the modalities of mimetic behaviors specific to humans, if they exist.*

Following this heuristic, I will quickly evoke some differences between animal and human cognitive capacities that could have a qualitative impact on imitation processes. From numerous studies in psychology, philosophy, neurology and ethology, two elements appear to play a crucial role in human behavior while being apparently out of the reach of non-human cognition.

First, humans are reflexive beings. To give a low level definition of reflexivity, it is the ability to take as an object of cognitive processing the cognitive processing themselves by creating new levels of cognitive processing. Some scholars claim that emergence of reflexive capacities can be traced through ontogeny through the study of the development of infant cognitive capacities (Zelazo et al. 1996) and the self-triggered loop that should be the elementary component of reflexive processes seems closely linked with the constitution of the self (Damasio 1999, Donald 1991, Mounoud 1995). Reflexivity allows us to think of others as we think of ourselves and of ourselves from other's point of view and thus develop our social skills. From the point of view of imitation processes, reflexivity makes all the difference since, as Eric Gans (1995) says, *"prehuman imitation is non-reflexive; the subject has no knowledge of itself as a self imitating another"*.

The second difference between animal and human cognitive capacities, closely related to reflexivity, is metacognition (Jacob 1998; Sperber 2000, Tomasello 1999), defined here as cognition about cognition. Whether animals have metacognitive capacities is still in debate in the scientific community. Some experiments seem to indicate that great apes and dolphins may have some rudimentary metacognitive capacities (Smith et al. 2003, Rendell and Whitehead 2001), but those are very limited. In particular, there is no evidence that animals can consider learning or imitation processes as object of cognition, and the fact that they do not teach tends to prove the contrary. Moreover, to our knowledge, there is no evidence that animals could voluntarily add a metacognitive level to solve a given problem, although some primates seem to be able to deal with chains of hierarchically organized behaviors (Byrne 1998). This means that animal metacognition, if it exists, is most probably constituted of rigid chains of process monitoring that could as well be hardwired, without requiring reflexivity to monitor their structure.

There is no space here to give more details about these two differences. But I will try to show now that taking them into account makes it possible to build a new class of models that may offer an answer to the problem of the endogenization of the distribution of imitation rules.

## *II Reflexivity of imitation rules*

To see what metacognition and reflexivity could change in the modeling of mimetic behavior, a more precise definition of what is an imitation rule is needed.

**II.1. Agents**

We will refer to an agent by the pronoun "it." In evolutionary games or in multi-agent modeling, agents are usually defined by a *n-tuple* $\tau$ of traits (age, color, opinion, behavior, rules of behavior, etc.) taken from a multi-dimensional set of traits *T*. In the following, we will place these traits into two categories: *modifiable traits* and *other traits*.

- ***Modifiable traits*** are those an agent can change voluntarily like for example a cooperative vs. defective behavior, the colors of clothes it wears, the political party an agent decides to vote for, the learning rules adopted for a given task, the chair where it wants to sit, etc. Most of the time, these changes take place on small time scales (in a day or so). The set of modifiable traits of a given agent *A* will be called its *strategy* and can be represented by an ordered *n*-tuple
  $s_A=(\tau_1,…, \tau_n) \in S$.
- ***Other traits*** are those that do not entirely depend on agent's will or are immutable. They depend on global dynamics and change generally on large time scales (months, years, lifetime), like social positions, payoffs, reputation, prestige, age, color of eyes, etc.

Agents are usually embedded in a social network and can learn some of the traits of the agents they interact with. For a given agent *A*, the neighborhood $\mathcal{N}_A$ will be defined as the set of all agents from which *A* can learn some traits.

Agents can then categorize their neighborhood into sub-neighborhoods on the basis of the learnable traits. For example, Boyd and Richerson (1985) consider different types of cultural transmission processes within sub-neighborhoods indexed to age and kinship: vertical transmission from parents to offspring, oblique from elders to younger, and horizontal among peers.

**II.2. Imitation rules**

Roughly speaking, imitation occurs when an agent decides to adopt a trait observed in one of its neighbors. For example, *A* may want to behave like *B*, wear the same clothes, or adopt its opinion. The most general definition for an imitation rule is thus a process that takes as input an agent *A* and its neighborhood and gives as output a modifiable trait *A* will try to copy from some of its neighbors.

From this broad definition, two distinct conceptions of imitation can be derived. Following René Girard (1961), *A* may want to be like *B* in some respect because *A* reads the values of things and actions in the eyes of others (def. 1). In that case, imitation precedes desire, and, from a formal point of view, it can be represented by some kind of conformist rule that governs desires (see Orléan 1985 for an example). The second conception of imitation is teleological and widely used in economics (Frank 2003) and multi-agent modeling (Conte & Paolucci 2001): *A* may want to be like *B* to a certain extent because from *A*'s point of view, *B* is a good model, i.e. *B* is more successful than *A* according to some criteria that *A* has adopted *independently* of knowing *B* (def. 2).

What we will propose here is in-between (def. 3): *A* may want to be like *B* to a certain extent because, from *A*'s point of view, *B* is a good model, i.e., *B* is more successful than *A* according to some criteria, that present some kind of *interdependence* with *B*'s criteria (for example because *A* has adopted them under *B*'s influence). Each definition is relevant in describing some kinds of social phenomena, but definition *3* is clearly missing in the modeling literature and could potentially bridge the two others. The clarification of this interdependence requires us to be precise about the above definition of "imitation rule" in a formal perspective.

> **Definition: *Imitation rule***
> Given an agent *A* and its neighborhood $\Gamma_A$, an imitation rule is a process that:
> 1. Assigns a value $v(B, \Gamma_A)$ belonging to an ordered set (the set of real numbers for example) to each agent *B* in $\Gamma_A$. $v$ will be called a valuation function.
> 2. Selects some traits to be copied from the best agents (according to the values given in 1) and defines the copying process.

For example, in payoff-biased imitation, the value assigned to each neighbor is its payoffs. The agent has then to infer which of the traits of the most successful neighbor is responsible for this success and try to copy it. In the case of a *conformist* rule, the value assigned to a neighbor is the size of the group it belongs to, and the traits to be copied are those of the largest group.

The valuation function here is subjective and dynamic, and it plays a role analogous to the utility function in game theory. Two agents can have different valuation functions and the diversity of valuation functions in a population expresses the diversity of points of view.

To synthesize the above definition, we can say that in step 1, potential models are selected whereas step 2 determines which of these potential models is going to influence the agent's behavior and how. It should be noted here that step 2 in the above description might introduce two sources of randomness:

1) $v(B, \Gamma_A)$ may assign the same best value to several agents with very different strategies. In that case if the agent has no particular preference on these strategies, it will pick one at random
2) Randomness can be explicitly coded in the rule. For example we can imagine a rule such that 1) everybody is a potential model, 2) the model is chosen with a probability proportional to its payoffs.

The goal here is not to be exhaustive, and this definition leaves a lot of things in the shadows, like for example the problem of inferring the relevant traits. Nevertheless, it is sufficient for our purpose, which is to propose a framework for thinking about an endogenous distribution of all kinds of decision-making rules evolving at the cultural time scale.

**II.3. Rules as modifiable traits**

The fact that human beings have reflexive and metacognitive abilities has some important qualitative consequences when it comes to modeling: individuals know to some extent that they are using rules for decision-making. This applies to imitation rules and, therefore, in formal models, imitation rules can be viewed as part of the strategy of the agent (as defined in II.1). They become modifiable traits. We should consequently study systems where imitation rules are modifiable traits by applying meta-rules.

This leads us to model agents as hierarchies of rules. A quick argument should convince us of the legitimacy of this representation. Considering the set of all rules an agent A is currently using, we can define a relation $\mathcal{R}$ in the following way. Let $r_1$ and $r_2$ be two rules used by A in its decision-making process, we will say that $r_2 \mathcal{R} r_1$ if the use of $r_2$ can change the way A uses $r_1$, which is equivalent to saying that $r_2$ acts upon $r_1$. For example, in the model of Orléan (1998, Table 1-c), the top-level rule $r_2$ (a kind of replicator dynamics) can change the rule $r_1$ used at level one (conformist rule or individual learning). We thus have $r_2 \mathcal{R} r_1$. On any set of rules, $\mathcal{R}$ defines a partial order that enables us to define a hierarchy of rules acting one on another.

For the sake of clarity, we will only consider in the following discussion agents such that $\mathcal{R}$ defines a total order on their set of rules. Moreover, as in most models already mentioned, agents will be defined by a unique type of behavior at the lower level controlled by a hierarchy of rules with a unique rule at each level. We will call such a hierarchy a metamimetic chain. As for models presented in table *1*, we can associate a given modifiable trait with a chain of imitation rules that controls its evolution.

Since our aim is to approach some aspects of human collective behaviors, we have to respect the currently accepted cognitive constraints. In particular, agents must have a bounded rationality. The consequence is that metamimetic chains have to be finite. This leads us to define the maximum length for such chains, called the cognitive bound of the agents ($c_B$): agents can modify the composition and the length of their metamimetic chain as long as the latter is inferior to $c_B$.

As for the top-level rules, we face two possibilities. Either we postulate a fixed exogenous rule, which is the option adopted by game theory and evolutionary game theory. In that case, top-level rules are interpreted as genetic determinants or fixed preferences (that nevertheless have to come from somewhere).

The other possibility is to consider that top-level rules are also modifiable traits: agents can act upon them. From our definition of $\mathscr{R}$, this is possible only if we assume that $r\mathscr{R}r$ when *r* is in a top-level position. We will now show that the above definition of imitation rule gives sense to this assumption and more precisely, that we can consider $\mathscr{R}$ as a reflexive relation, i.e., $r_k \mathscr{R} r_k$ for all *k*.

## III. Metamimetic dynamics and endogenization of meta-choices

### III.1 Agents as metamimetic chains

In this section, I will explore some possible dynamics in a population of agents described in terms of chains of imitation rules, the relation $\mathscr{R}$ being reflexive. For reasons of clarity and because they have been extensively studied, we will not evoke other characteristics of decision-making processes like anticipation, learning or memory, although they plays an important part and should be taken into account in future work. The goal here is to identify some remarkable properties of metamimetic dynamics.

Consider a population of agents defined by metamimetic chains that can deal with a maximum of $c_B$ meta-levels (bounded rationality). Consider that agents have to choose a behavior $r_0$ among several possibilities (like *C or D*). This behavior has some impact on their environment, like, for example, an influence on their material payoffs or on the densities of the different kinds of behaviors in their neighborhood. Assume that agents can change their behavior with metamimetic chains composed of rules taken in a set *R*.

Agents will then be defined by a set of modifiable traits[iii] *($r_0,r_1,.. r_n$)* with the constraint $n \leq c_B$, where *$r_0$* is a behavior and *$r_j \in R$* for *j>0* (figure 1).

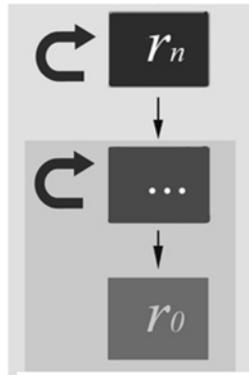

**Figure 1: Schematic representation of a metamimetic chain.**

With this representation, the activity of an agent will consist in the ongoing checking of the coherence of its hierarchy of decision-making processes (non contradiction between rules), focusing its attention on different levels at different moments. For example, if the agent focuses its attention to level *k,* since for *$j \leq k$* $r_k \mathcal{R} r_j$ (*$\mathcal{R}$ is reflexive)* the question it will ask itself will be: "Is the strategy *($r_0,r_1,.. r_k$)* the best one from the point of view of *$r_k$* ?". If this is not the case, the agent will try to change its strategy for a better one. Except for the *status quo* case we can identify two categories of possible outcomes in the revising of *($r_0,r_1,.. r_k$):*

- 1) Some elements in *($r_0,r_1,.. r_{k-1}$)* are modified but the initial rule *$r_k$* is still part of the strategy.
- 2) *$r_k$* has been modified.

We will now examine these two possibilities. To fix ideas, in our examples, we will consider agents with two different action opportunities, *C* and *D,* which have two different material consequences (monetary payoffs for example). As for the set of rules *R,* we will take the most common rules in the modeling literature: imitation of the most successful agent in terms of material consequences (*payoffs-biased* or *maxi* rule for short) and imitation of the most common traits (*conformist* rule): *R={maxi, conf}.* It should be emphasized that payoffs here should not be interpreted as a direct mapping of what could be the utility function of agents. The analog to utility functions is the hierarchical set of valuation functions of a metamimetic chain.

### III.2. Endogenous variation of length in metamimetic chains

The standard way of activating a rule at level *n* is to change the trait of the level below. For example in Orléan 1998, an agent *A* with *$s_A$=($r_0,r_1,r_2$)* can decide to change its behavior *$r_0$* from *H* to *L* when it takes the point of view of its rule *$r_1$* (for example the conformist rule) (cf. figure 2). But it can also change its rule $r_1$ when focusing its attention on the rule $r_2$, i.e. *maxi* if it happens that $r_1$ is not the most successful rule in its neighborhood (in this article, a random sample in the population).

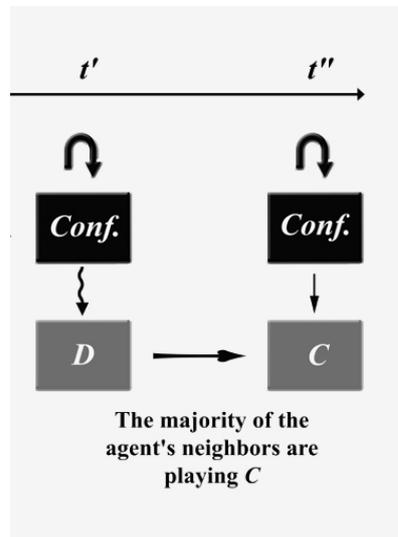

**Figure 2: Update of an intermediary modifiable trait.** A conformist agent *A* observes that the majority of agents is playing *C*, and decides to update its behavior to *C (modifiable trait of level 0).*

When we consider agents that can monitor the complexity of their strategy by adapting the composition and the length of their metamimetic chains, this standard way of using rules should be generalized. Let's begin with a simple example.

Assume that after the observation of a conformist agent *B*, *A* with $s_A=(D, maxi)$ comes to the conclusion that to maximize its payoffs, the best thing is to do like *B*. In that case, if its cognitive bound is large enough, the most rational behavior is to adopt the strategy $s_A'=(D, conformist, maxi)$, and act as a conformist as long as this rule is adaptive from the point of view of the *maxi-rule* (cf. figure 3). Doing this, it will eventually be able to come back to its initial strategy ( * , Maxi) in a subsequent imitation.

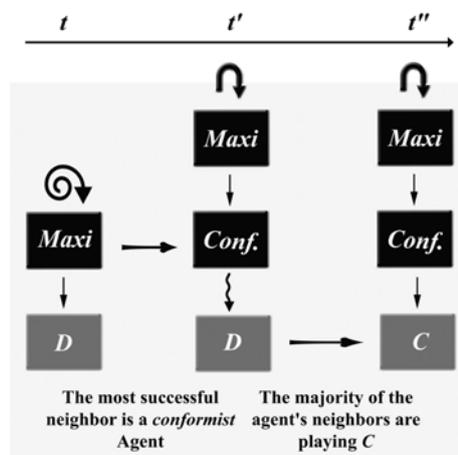

**Figure 3: Endogenous variation in the length of metamimetic chains.** At time *t*, a *Maxi* agent *A* has a *conformist* neighbor that is more successful than all agents in ᴀ. *A* will then adopt the conformist rule at its first meta-level, keeping in mind that it is only a means for maximizing its payoffs (second meta-level). Thereafter, it might be that according to this conformist rule, the current behavior is not the best one and has to be changed.

In that case, the complexity of *B's* strategy and the cognitive bound of *A* enable *A* to keep in mind its initial rule $r_1$. *B's* strategy is a temporary means for achieving the goals defined by $r_1$. This kind of transition enables the agent to change endogenously the length of its metamimetic chain. We do this kind of mental operation every day every time we decide that the realization of a goal *G'* is the best way to achieve a goal *G*.
This simple example shows how imitation can lead to endogenous variation in the length of metamimetic chains.

### III.3. Reflexive update at the limit of the cognitive bound

The metacognitive skill that allows agents to change the structure of their strategy for a more adaptive one has its inevitable counterpart, implied by their bounded cognitive capacities. Take for example an agent *A=(D,maxi)* with a cognitive bound of *1* . Assume *A* finds out that one of its *conformist* neighbors has higher payoffs than any other neighbor, *A* included. To try to be as successful as *B, A* will have no other solution than to *become conformist* (figure 4).

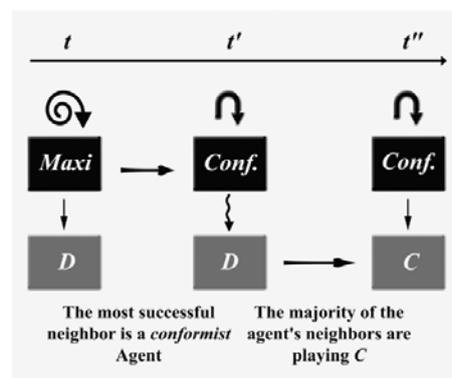

**Figure 4:** *Reflexive update at the limit of the cognitive bound.* At time *t,* a *Maxi* agent *A* has a *conformist* neighbor that is strictly more successful than all other neighbors. Consequently, *A* adopts the *conformist* rule. Thereafter, it might be that according to this rule, the current behavior is not the best one, and has to be changed.

A metamimetic agent will encounter this kind of situation each time the strategy of its model is so complex that it cannot manage both new goals and old ones. In such cases, and when the updating rule is the top-level one, *A* will have to either stop attempting to improve its strategy or to change its top-level rule. The first option is the status quo, while the second one radically changes the decision making process of the agent and consequently the dynamics of the social cognition. Should the slightest proportion of transitions belong to the second type, a metadynamics on social cognition will emerge.

The second type of transition is formally well defined in our framework: since imitation rules are modifiable traits, it might happen that the trait to be modified in an imitation process defined by a rule *r* is the rule *r* itself. In that case, an agent *A* will change reflexively its rule *r* because, **from the point of view of *r*, *r* is not the most adaptive rule**. In such situation, we can say that *r* is not self-coherent in *A's* environment: it prescribes actions that are in conflict with the continued possession of *r*.

This updating process gives sense to the fact that rules for imitation can be their own metarule. We will say that an imitation rule can *update reflexively by acting on itself as a modifiable trait*.

We can now comment on the reflexive updates idea. People certainly don't have a wired cognitive bound that obliges them to do such clear-cut transitions in their decision processes. Nevertheless, it often happens that an activity that was first considered as a means becomes an end in itself. For example it might come to finally take up all our time; we might forget its primary purpose or we might simply come to like the new activity more than the previous one. In these examples, the important feature is that the new goals do not come from nowhere but are related to the old ones to some extent. The reflexive mimetic update presented above is a stylized representation of these kinds of transitions when they are triggered by the observation of others.

Viewed at the population level, these kinds of updates define an endogenous dynamics on top-level rules that is worth studying.

## IV The metamimetic game

**IV.1 Definition**

The characteristics of imitation rules introduced in the preceding section (figure 5) suggest a new class of models for the modeling of mimetic dynamics.

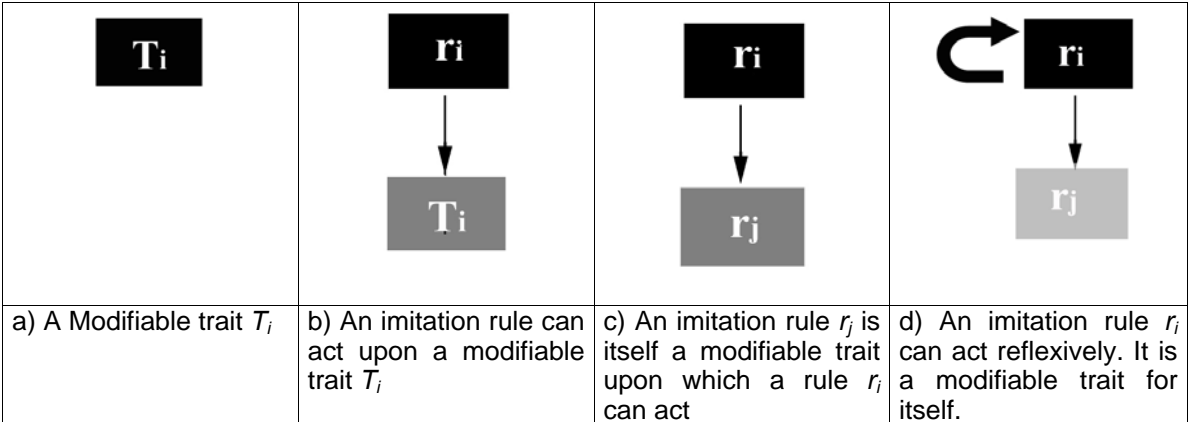

| a) A Modifiable trait $T_i$ | b) An imitation rule can act upon a modifiable trait $T_i$ | c) An imitation rule $r_j$ is itself a modifiable trait upon which a rule $r_i$ can act | d) An imitation rule $r_i$ can act reflexively. It is a modifiable trait for itself. |

**Figure 5:** When imitation rules are modifiable traits, they can be modified by other rules (c) and can be modifiable traits for themselves (d).

**Definition: *Metamimetic game***

Let B be a set of behaviors and R a set of imitation rules, a *metamimetic game* $G=\{N,\Gamma,R,B,C_B\}$ is an N-players game where each agent A is characterized by a metamimetic chain $s_A=(r_0,r_1,..\ r_k,)$ with $r_0 \in B$ and $r_j \in R$ for $j>0$.

*Moreover, the three following conditions should be satisfied:*

*C-I - **Bounded rationality**: the number k of meta-levels in metamimetic chains is finite and bounded for each agent by its cognitive bound $c_B$ ($k \leq c_B$).*

*C-II - **Metacognition**: at all levels in a metamimetic chain, imitation rules are modifiable traits.*

*C-III – **Reflexivity**: imitation rules can update reflexively changing the length of the metamimetic chain in the limit of the cognitive bound of agents. When the cognitive bound is reached, imitation rules may update themselves.*

Studying metamimetic games will thus consist in the study of the evolution in the length and composition of these chains, leading to the emergence of structures at the intra and inter-individual levels.

The main difference with other kinds of games from game theory and evolutionary game theory is that in metamimetic games, there is an endogenous dynamics on the distribution of rules and metarule whatever the dimension of the rule space (even the top level is not necessarily a singleton). There is not enough space here, but by writing the master equation of this kind of games, we can demonstrate that contrary to other games of imitation like replication by imitation (Weibull 1995) metamimetic games are not in general reducible to some standard dynamics like the replicator dynamics. In fact, the discrete replicator dynamics (Hofbauer & Sigmund 1988) is reducible to a particular case of metamimetic dynamics where the set of possible top-level rules is a singleton (see a brief note in the appendix).

We will now make intuitive these differences on a minimal example. This will allow us to introduce the main concepts related to these games.

### IV.2. An example of a minimal metamimetic game

Consider the following metamimetic game:
- Two agents *A* and *B* with a cognitive bound of *1 and no memory,*
- *Each agent is in the other's neighborhood ($\Gamma$ is a fully connected graph),*
- A binary choice for action: *C* and *D,*
- A binary choice for imitation rules: *maxi* and *conformist,*

- *The game is symmetric and when C plays against D, D gives always higher payoffs than C (think of a prisoner dilemma for example),*
- The game is repeated and at each period, agents change their strategy *simultaneously* according to their rule.

The precise definition of the rules is the following:
- *Maxi:* if your neighbor has higher payoffs than you, copy its rule and use it to update your behavior.
- *Conformist:* If your strategy is different form the one of your neighbor, copy its rule and then use it to update your behavior.

The state of the game is thus given by the behavior and the imitation rule of each agent like for example: *s=[$s_A$:(C, maxi); $s_B$: (D, conf.)]*.

There are only 16 possible states and the metamimetic dynamics defines a Markov chain on this set (figure 6).

For example if the initial state is *s=[(C, maxi); (D, conf.)]*, after one period, A will become conformist because B is the most successful agent, and will change its behavior from *C* to *D* to adopt B's behavior. B will become *maxi* to be like *A* and will keep on playing *D* because it is the most successful action. In the second period, both agents will have the same behavior and consequently will have the same payoffs. Then only A will change its strategy to be like B and both will end with the strategy *(D, maxi)*. The final state of the game will be *s"=[(D, maxi); (D, maxi)]*.

We will say that *s"* is reachable from both states *s=[(C, maxi); (D, conf.)]* and *s'=[(D, conf); (D, maxi.)]*. More generally, we will say that a state *s'* is reachable from a state s if and only if a system starting in state *s* can reach the state s' after a finite number of mimetic transitions.

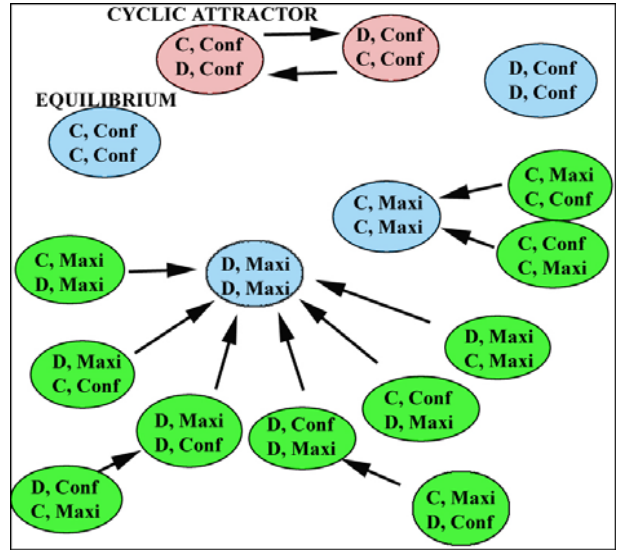

**Figure 6: The dynamics in a minimal metamimetic game.** Each arrow represents the mimetic transition that the current state requires (here with probability 1 everywhere).

We will now introduce the main concepts of stability for metamimetic games. For the sake of clarity, these concepts will be based on the minimal example presented in this section ($C_B$=1, discrete set of rules). However, these concepts generalize to situations where the cognitive bound is higher than one and are in no way dependant on the discreteness of the set of rules. For example, in the same way game theorists consider mixed strategies, one can imagine a model representing valuation functions as the weighted sum of several valuation functions corresponding to discrete rules. In that case, an imitation rule would be defined by the way the agent changes the weights associated with different discrete imitation rules, which do not even have to be a pure copying process but a simple 'move toward'.

### IV.3. Metamimetic equilibrium

Two types of remarkable subsets of states in figure *6* should be highlighted, equilibriums and attractors. They are the remarkable subsets of the underlying Markov chain.

**Definitions:**

Let *G={N,Γ,R,B,$C_B$}* be a metamimetic game with *$C_B$=1 and Ω the set of all possible states of the game*.

- A set of states *Σ=($s^1$,…, $s^m$) is a **metamimetic attractor** if and only if*

$\forall (s,s') \in \Omega \times \Sigma$; *s' is reachable from s $\Leftrightarrow$ s'$\in \Sigma$*

- A state $s=(s_1,…,s_n)$ is a **metamimetic equilibrium** if and only if $\{s\}$ is a metamimetic attractor. In case of our minimal example ($C_B=1$) this can be expressed by the following condition:

$$\forall i \in \{1,…,N\}, k \in \Gamma_i, s_i=(r_0^i, r_1^i) \neq s_j=(r_0^j, r_1^j) \Rightarrow v^i(j, \Gamma_i) \leq v^i(i, \Gamma_i)$$

"no agent can find itself better when it imagines itself in the place of one of its neighbors"

where $v^i$ is the valuation function associated with the imitation rule $r_1^i$. Here the value $v_i(j, \Gamma_i)$ can be understood as a counterfactual, it is the well-being that $i$ can imagine feeling while being put at the place of $j$. In this sense, a metamimetic equilibrium will also be called a *counterfactually stable state*. This definition can be extended to any value of $C_B$.

This very simple example has the advantage of making clearly visible the main property of metamimetic games as illustrated in figure *6*: the Markov chain that defines the dynamics is a property of the set of strategies considered, not a dynamics that would be given apart from this set, like for example a *maxi* rule or a replicator dynamics applied to a set of strategies. We will further discuss the choice of this set in section *V.1*. This property is independent of the size of the set of rules and is the key for the expression of a multiplicity of agents' viewpoints. At this stage, we can clearly see the link with the notion of *operational closure*: dynamics on imitation rules for a given state of the game are the product of the distribution of imitation rules.

Moreover, the goals of an agent at a given moment, defined as the valuation function of its unique imitation rule, are the outcomes of its ongoing interaction with its environment. We thus shifted from a perspective where goals were an unchanging property of agents (like the standard maximization of predetermined payoffs in economy) to a perspective where goals are to some extent chosen by an agent during its life under the influence of its social interactions. The distribution of goals in a population should then be understood as the expression of the self-coherence of these goals in the social network and not the expression of the fitness of these goals relatively to criteria whose distribution exists prior to the social activity.

To return to the comparison with replicator dynamics, for the latter, the meaning of a transition from a global state *s* to a global state *s'* should be interpreted in terms of a semantics that is external to the system, the fitness function (what is good, what is bad). The different states of the system have no meaning other than the one assigned from the outside. On the contrary in metamimetic games, the meaning of a transition from *s* to *s'* has to be found in *s* itself, it is the expression of the content of imitation rules.

We can sum up these remarks with the following proposition:

**Proposition:** *Every metamimetic game $G=\{N,\Gamma,R,B,C_B\}$ can be associated with a unique matrix $P^0$ that determines the Markov process representing the internal dynamics of the game. $P^0$ defines the metadynamics of the social cognition process.*

The proof of this proposition is straightforward since for each configuration of the social network, agents' imitation rules define locally the possible transitions (I remark in passing that transitions do not have to be deterministic since for a given agent there could be several models with distinct strategies).

The internal dynamics of a metamimetic game defines some particular distributions corresponding to eigenvectors of $P^0$. This is a first spontaneous selection among all possible states. In our example, from the sixteen possible states, only six are attractors. As we will see, this selection is sharpened in presence of perturbations.

### IV.4. The role of perturbations

In the preceding section, agents were supposed to be mind readers: they knew perfectly well the strategies of their neighbors. This of course does not pretend to reflect reality since in real setting, people have to infer rules, behaviors and other cognitive components of the decision making process of their neighbors. These inferences about what others think and do are noisy. Moreover agents do not always do what they intend to do. Consequently, there are errors all along the decision-making process due to false perceptions, misunderstandings and mistakes. Following Young (1993, 2001), a more realistic approach would be to suppose that there are some mistakes that constantly perturb the social dynamics. We have then to study metamimetic dynamics in the framework of stochastic game theory (Foster & Young 1990).

For example, as a first approximation we can assume that all mistakes in copying neighbors' strategies are possible and time-independent and consider that with a probability $1-\varepsilon_r$ (resp. $1-\varepsilon_a$) the agents choose the correct imitation rule (resp. behavior) but with a probability $\varepsilon_r$ (resp. $\varepsilon_a$) the rule chosen (resp. the behavior) is picked at random in the set $R$ (resp. $B$). We obtain a disturbed Markov process $P^\varepsilon$ that has a unique stationary distribution. When the perturbation is small ($||P^\varepsilon - P^0||_\infty << 1$) this stationary distribution is concentrated around a particular subset of attractors of the process defined by $P^0$: the stochastically stable set (Foster & Young 1990). To highlight the fact that this Markov process $P^0$ represents endogenous mimetic dynamics, we will call this set the stochastically counterfactually stable set (*SCSS*). In our example, it is straightforward to show that the only state in the *SCSS* is the state *[(D,maxi); (D,maxi)]* (figure 7 & 8).

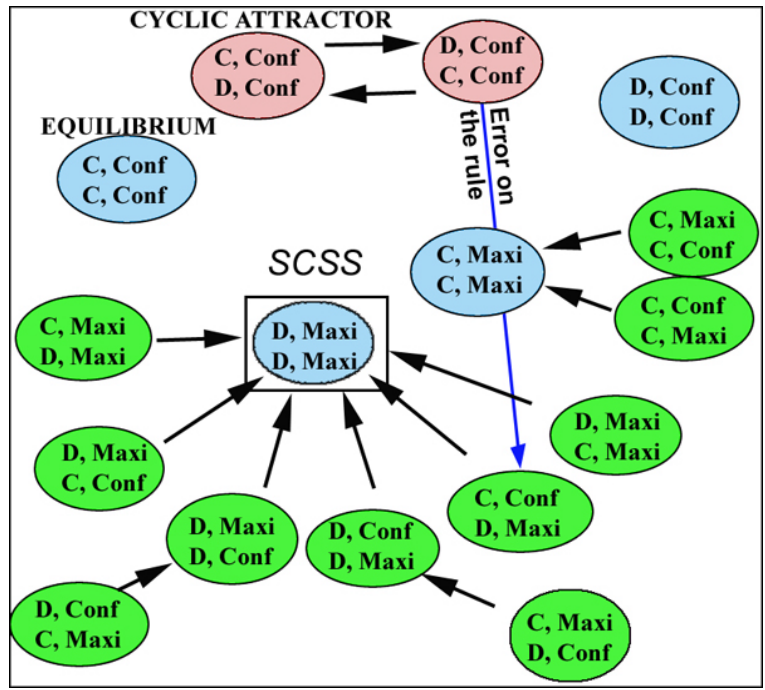

**Figure 7: Minimal example of a noisy metamimetic game.** An error of one of the agents can lead the system toward a new metamimetic attractor. When the system is constantly disturbed, the only state in the SCSS is the state where all agents are (*D,maxi*).

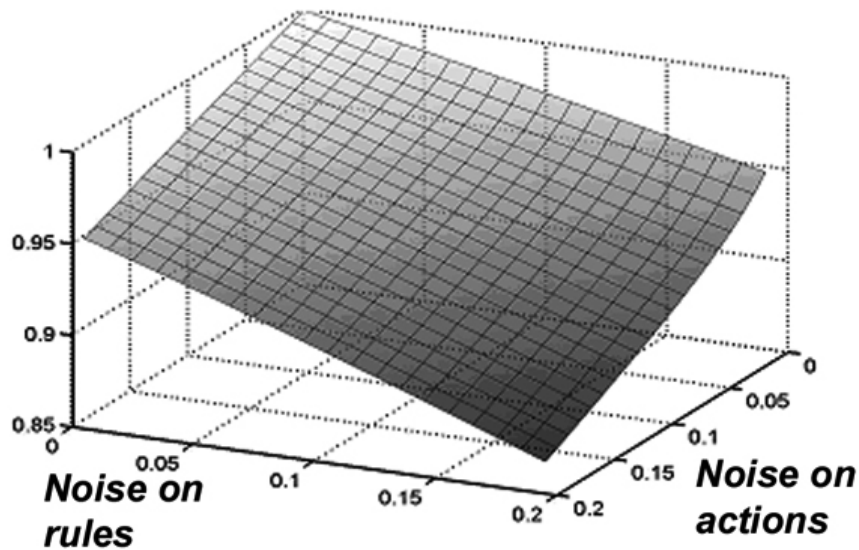

**Figure 8: Frequency of the SCSS at the attractor.** In our minimal example, the proportion of the SCSS in the limit distribution when ($\varepsilon_r$, $\varepsilon_a$) tends to zero quickly converges to *1*. Here the proportions have been plotted for *0,01<$\varepsilon_i$<0,2*. (It should be noticed that since all states are taken with an equal probability in case of mistake, here half then errors are corrected. The real level of noise is then $\varepsilon_i/2$)

The simplicity of our example does not allow us to show all the characteristics of these games in noisy settings. The most important feature here is that the coupling between the internal dynamics and the perturbations operates a second selection on the states privileged by the internal dynamics, the limit distribution being independent of the initial conditions.

There is not enough space here to expose the role of perturbations in detail; I will only mention that usually, the SCSS does depend on the structure of the perturbation which here is given by the relation between $\varepsilon_a$ and $\varepsilon_r$. Formalizing this dependence is perhaps one of the principal contributions of this approach. In those systems, we have two sources of information: the internal dynamics and the structure of noise in the environment. The relation between these two is an example of structural coupling (Varela 1979).

But the main aspect this simple example highlights is that dynamics of metamimetic games are not optimization of criteria that are arbitrarily assigned to agents and exist prior to their activity. The question here is not to find the "best" strategy, which would first require that the modeler ask herself which definition of best is the best. The question is to find the states such that each agent, through its interactions with its social networks, has found an identity and a strategy that is self-coherent given its social environment. The question is then to find the social configurations that are maximally stochastically counterfactually stable. This suggests us to study social systems as autonomous systems evolving under processes of differentiation. This differentiation takes place among a multiplicity of possible criteria in a process of cultural co-evolution.

## *V. How to choose the set of imitation rules?*

### V.1. Perceptions and computations

Before concluding, I would like to highlight the shift of perspective about social systems modeling that the current approach could bring about. Contrary to standard games where the modeler has first to choose a set of strategies and then a dynamics on this set, in metamimetic games, the set of strategies is sufficient to define the dynamics. For this reason, we must pay a particular attention to this set.

Following Baldwin (1897), who wrote that imitation is a means of selecting stimuli in the environment, it is interesting to think of imitation rules in terms of selective attention: an agent is particularly sensitive to one dimension of its perceptive space and builds from the stimuli detected along this dimension a function that it will use to select the appropriate trait to copy. This suggests that we should not define the set of imitation rules as a list, as is usually done, but as a set generated by some cognitive operators: operators for the selection of a particular dimension in the stimuli space and operators for computation on this selection (figure 9).

To choose the appropriate set the modeler must ask herself, "What can an agent perceive in the situation considered?" and "What kind of operations can an agent do on these perceptions?" For example, the fact that an agent is able to act as a conformist means that it can focus its attention on frequencies of traits in a population and find the most frequent trait. The fact that an agent is able to act like a payoffs-maximizer means that it can focus its attention on payoffs and can at least compute the maximum of two scores.

**V.2 Regularities of the set of rules**

This approach suggests that an important part of the work is to use cognitive sciences research to find regularities of such sets of rules in order to build them in a generative way. The first regularity we can propose as a first approach has been evoked by Gabriel Tarde (1890) in the forewords of the second edition of his *"Laws of Imitation":* we can reasonably think that there is no social system where an imitation rule is present without its counterpart, the one that prescribes exactly the contrary. Indeed, if an agent can imagine a rule, there is no reason to think that it cannot imagine its contrary. From a formal point of view this means that if an agent can find the maximum between two numerical values, it can also find the minimum, which is formally equivalent to saying that there is a cognitive operator that multiplies numerical values by *–1.*

If we accept this property, the minimal set of rules that contains the *maxi* rule and the conformist rule (figure 9) would be in fact a set with four rules {*maxi, mini,* conformist, non-conformist}, where the basic percepts are densities of traits and payoffs, and the operators *inverse* (that multiplies numerical values by *–1*) and *max* (that find the max between two numerical values). This means that if one of these rules plays a minor role in the model considered, it should be because it has been selected against by the internal dynamics of the system and not because the modeler has decided against this rule a priori. This is possible precisely because the distribution of imitation rules is endogenous once you choose a set.

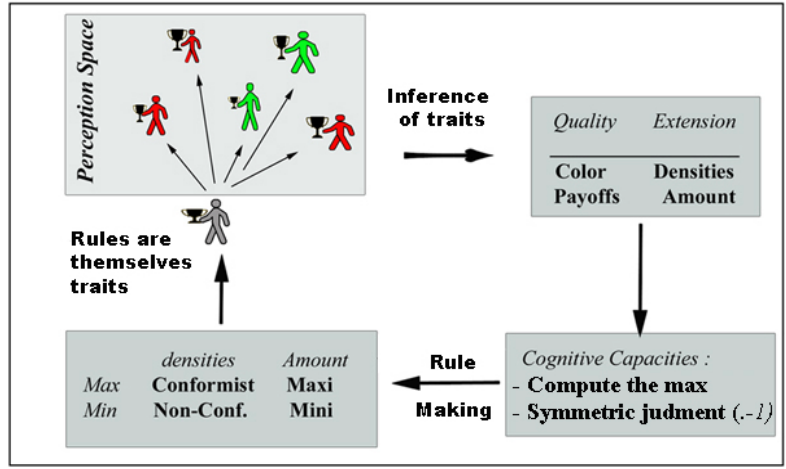

**Figure 9: Our approach suggests defining the set of imitation rules in a generative way.** Agents are embedded in their environment from which they infer some traits like colors and payoffs. Then, they do some computations on the inferred distributions of these traits: computation of the densities, computation of the maximum of payoffs, computation of the minimum, etc. These computations are used as a basis for building imitation rules that are themselves traits agents can try to infer.

**V.3. Shifts in the phase space**

These remarks raise a fundamental question: How can new categories of traits, new behavioral options, new dimensions emerge in the perceptive space that will bring the metamimetic dynamics from a Markov chain defined by $P^0$ to an other defined *by P'$^0$*? Is it possible for example to imagine a population of agents that, at a certain moment in a simulation, become sensitive to the number of imitators per agents so that they begin to imagine rules like "imitate the one that is the most imitated", (i.e. prestigious agents)? Is it possible to imagine how a population of *conformist/non-conformist* agents who care only about densities of traits can invent such a thing as money, which brings about new couples of rules like *maxi/mini*? Such transitions reflect a discontinuous shift in the phase space of the whole systems and their study is highly relevant to the study of social systems dynamics. Indeed, it is highly probable that social systems are constantly shifting from phase spaces in that way as a consequence of innovations. Understanding these kinds of shifts is certainly one of the challenges of social systems modeling.

It should be highlighted however that our model in its present formulation, couldn't simulate such shift, since it cannot simulate strong emergence. As for all computational methods, like genetic algorithms for example, we can only approximate the phenomenon of emergence, defining from the beginning the whole space of possible options, starting with a strict subset of options and enlarging this subset with random explorations (noise). Actually, there is little hope to be able to do better than this approximation in the framework of social simulations. Indeed, if a rule, a category or a behavioral option has to be present at a given moment in a simulation, it will necessarily correspond to an expression written in the language *L* of the program. Considering as the set of all possible options/categories/rules the (possibly infinite) closed set of all valid expressions in *L*, the program will never be able to get out of this set. Consequently, emergence of brand new options that aren't conceivable from the beginning is not possible in computational studies. The evaluation of what we miss by modeling open phase spaces with closed spaces is still an open question. Can we approach as close as desired the qualitative properties of open spaces with simulations on closed sets, in the same way we can approach irrational numbers with rational numbers?

## *VI Conclusions*

As a heuristic for the modeling of human social systems, several scientists have proposed focusing on models that include human specific cognitive capacities. The justification for this is that only such models should be able to explain the huge gap in the complexity of social structures between animal and human societies. Following this heuristic, we proposed including some consequences of reflexivity and metacognition in social systems modeling, i.e. the fact that human beings know to some extent that they are using rules for decision-making and can monitor their use.

When associated with reflections about human imitation, this remark led us to propose a formal framework for modeling mimetic dynamics within social systems: *metamimetic games*. In those games, agents imitate in accordance with their preferences, as in most models that deal with imitation, but also form their preferences through imitation. Those games have two remarkable properties:
1) Imitation rules can be their own meta-rules. Thus, we escape the regress problem that threatens as soon as the question of grounding the choice of strategies is evoked;

2) There is an endogenous metadynamics of imitation rules: the distribution on imitation rules is the product of the dynamics it defines.

With a very elementary example of such a game, we introduced the corresponding concepts of attractors and equilibriums: counterfactually stable states. In a noisy setting, we worked in the framework of stochastic game theory (Foster & Young 1990) and proposed the concept of *stochastically counterfactually stable states*.

Unlike other games, the main question in the study of metamimetic games is not to find the "best" strategy. This would require indeed that the modeler knows which definition of "best" is the best. Rather, the issue is the study of a phenomenon of differentiation in a population along a process of cultural co-evolution that takes place among a set of possible motivations. This consists in identifying the (stochastically) counterfactually stable states that are the states toward which the dynamics converges. Those are states such that the social positions of agents are maximally coherent[iv] from *their own* point of view given what they can do.

This is only the first outline of metamimetic games. As already mentioned, future work will have to develop models adapted to more specific situations and study the links between metamimetic dynamics and other components of human cognition such as perception, inference, learning, and eduction. In particular, we could question the concepts of rule revision and reflexive update from the perspective of new theories about the dual nature of decision-making processes (Kahnemann 2002, Lieberman et al. 2002, Stanovich and West 2000). Finally, it is also desirable to include forward-looking and imitation in a single framework, since it is well known that people make decisions after what they have experienced, what they can see around, and what they can imagine. In such a framework, mimetically inspired strategies might be considered as anchors for local search in the strategies' space through variation and eductive projections.

An other interesting issue is the study of the interactions between mimetic dynamics and other kinds of social dynamics. There always exist some prerequisites for participation in a social activity. For example, to participate in an economic activity, one must be credit worthy. These kinds of prerequisites induce a dynamics in the population – often modeled as a replicator dynamics - that is superimposed on the metamimetic dynamics. We can expect that the understanding of the mutual influence between these different dynamics will be very instructive in understanding the complexity of human social systems.

**Acknowledgments**

The author would like to thank T.K. Ahn, Paul Bourgine, Jean-Pierre Dupuy, Jean Petitot, Lucien Scubla, Gérard Weisbuch for helpful discussions and comments. Special thanks to my referees, for their kind attention and bright suggestions.

This paper was supported by the ACI Complex Systems in SHS "Réseaux Coalitions et Marchés" and the Polytechnique School, CREA, Paris (http://www.polytechnique.edu).

# Appendix: Metamimetic dynamics are not reducible to standard replicator dynamics

In the literature of social modeling, one of the most frequent mimetic dynamics is a replication by imitation indexed on payoffs. This dynamics appears to be reducible to the replicator dynamics (Weibull 1995). We will show here that this is not the case for metamimetic dynamics in general and highlight the structural difference between metamimetic dynamics and dynamics similar to the replicator dynamics.

To begin with, the most general equation we can write about metamimetic dynamics is a balance equation. If we write $x^t_c$ the proportion of metamimetic chains equal to $c$ at time $t$ and $p^t_{cc'}$ the probability for an agent with a strategy $c$ to switch for a strategy $c'$ at time t, we have:

$$\Delta x^t_c = x^{t+1}_c - x^t_c = \Sigma_{c'} x^t_{c'} \cdot p^t_{c'c} - x^t_c \Sigma_{c'} p^t_{cc'} \quad \text{(Equ. 1)}$$

We now have to interpret this equation in our framework and particularly the terms $p^t_{cc'}$. In metamimetic games, when individuals change their strategy, they try to copy the best strategy in their neighborhood, the term "best" being relative to their metarule. Consequently, the terms $p_{cc'}{}^t$ will have various functional forms as soon as there will be more than one metarule available, whereas the replicator dynamics presents only one functional form.

Indeed, it is straightforward to propose a metamimetic game that follows a replicator dynamics by considering a game with only one type of metarule. The standard form of discrete replicator dynamics for a population of strategies $(1,..,n)$ with proportions $\sigma(t)=(x_1^t,...,x_n^t)$ can be written (Hofbauer and Sigmund 1988):

$$x_i^{t+1} = \frac{a + f_i(\sigma_t)}{a + \hat{f}(t)} x_i^t$$

with $\hat{f}(t) = \sum_{i=1}^{n} x_i^t f_i(\sigma_t)$

It can be rewritten:

$$\Delta x_i^t = \left( \frac{a + f_i(\sigma_t)}{a + \hat{f}(t)} - 1 \right) x_i^t$$

Consider now a metamimetic game on a complete graph with a single rule *r* and a set of behaviors *(1,..,n)*. The metamimetic chains can then be named after the associated behavior. Assume that the rule *r* follows a stochastic metamimetic rule: an agent *A* will imitate a neighbor *A'* with a behavior *j* with a probability proportional to $\frac{a+f_j(\sigma_i)}{a+\hat{f}(t)}$. Since each neighborhood contains the whole population (the graph is complete), we then have:

$$\forall i,j : p_{ij}^t = \frac{a+f_j(p_j^t)}{a+\hat{f}(t)} p_j^t .$$

We replace this relation in Equ. 1:

$$\Delta x_i^t = \Sigma_j x_j^t \cdot p_{ji}^t - x_i^t \Sigma_j p_{ij}^t$$

$$\Delta x_i^t = -x_i^t \cdot \sum_{j \neq i} \frac{a+f_j(x_j^t)}{a+\hat{f}(t)} x_j^t + \sum_{j \neq i} x_j^t \frac{a+f_i(x_i^t)}{a+\hat{f}(t)} x_i^t$$

$$\Delta x_i^t = -x_i^t \cdot \frac{1}{a+\hat{f}(t)} \sum_j x_j^t (a+f_j(x_j^t)) + \frac{a+f_i(x_i^t)}{a+\hat{f}(t)} x_i^t \sum_j x_j^t$$

$$\Delta x_i^t = \left( \frac{a+f_i(x_i^t)}{a+\hat{f}(t)} - 1 \right) x_i^t$$

The discrete replicator dynamics translated in terms of metamimetic dynamics is thus equivalent to the particular case of a metamimetic game with a single meta-rule that can be formulated by "imitate neighbors at random with a probability proportional to their fitness".

Consider now a straight generalization of the game $G=\{N,\Gamma,R,B,C_B\}$ proposed in *IV.2*, with *N>1* and $\Gamma$ being a complete graph. In this setting, all the agents with the same strategy *c* have the same update behavior defined by a set of transition probabilities $p_{cc'}^t$ *(probability at time t for an agent to switch from c to c' )*. There are four possible metamimetic chains and we can write the matrix transition $(p_{c'c}^t)$:

| $p^t_{c'c}$ | (C,conf) | (D,conf) | (C,maxi) | (D,maxi) |
|---|---|---|---|---|
| **(C,conf) or (D,conf)** | Equals 1 if conf. agents are in majority and C is the most common behavior, 0 otherwise. | Equals 1 if conf. agents are in majority and D is the most common behavior, 0 otherwise. | Equals 1 if maxi agents are in majority and C is the most successful behavior, 0 otherwise | Equals 1 if maxi agents are in majority and D is the most successful behavior, 0 otherwise |
| **(C, maxi) or (D, maxi)** | Equals 1 if conf. agents are the richest and C is the most common behavior, 0 otherwise. | Equals 1 if conf. agents are the richest and D is the most common behavior, 0 otherwise. | Equals 1 if maxi agents are the richest and C is the most successful behavior, 0 otherwise. | Equals 1 if maxi agents are the richest and D is the most successful behavior, 0 otherwise. |

We can see that this game is not reducible to a replicator dynamics, since the terms $p^t_{c'c}$ have several functional forms where both the notion of "most successful" and "most common" come into play. It is also easy to identify all metamimetic equilibriums associated to this very simple game. They are of three types:

1. All the population is composed of maxi agents that uniformly defect or cooperate.
2. All the population is composed of conformist agents that uniformly defect or cooperate.
3. The population is composed of a fraction $p_{conf} \geq 0.5$ of conformist agents that uniformly defect or cooperate and a fraction $1 - p_{conf}$ of maxi agents that uniformly defect or cooperate, the latter option being possible only if all conformist agents also cooperate.

---

[i] It should be clear that the definitions given here for imitation rules are only operational definitions borrowed from literature. They are not meant to exhaust the underlying social manifestations of imitation. Moreover, the precise definition of an imitation rule in an operational perspective is an aspect of social modeling that is complementary to the current topic, i.e., how the distribution of these different kinds of rules might evolve under social cognition.

[ii] « *Pour élaborer une science de l'homme, il faut comparer l'imitation humaine avec le mimétisme animal, préciser les modalités proprement humaines des comportements mimétiques si elles existent* » Girard, 1978.

[iii] Also it could be interesting to consider different sets of rules depending on the agents – as in Selten & Ostmann 2001 - and the cognitive level, but this would be superfluous given the present purpose.

[iv] The term coherent should be understood before all from the modeler's perspective. Agents themselves are not looking for coherence but are simply applying their rules. A rule is self-coherent if its application does not tend to change the rule itself.